\documentclass{aa}

\newcommand{\Mup}[0]{$M^{max}_{COWD}$~}
\newcommand{\MCOcore}[0]{$M_{CO}$~}

\newcommand{\Mupstar}[0]{$M^{min}_{CCSN}$~}
\newcommand{\msdu}[0]{$M^{min}_{SDU}$~}

\usepackage{amssymb, amsmath}
\usepackage{graphicx}
\usepackage{txfonts}
\usepackage{natbib,twoopt}
\usepackage[usenames,dvipsnames]{color}
\usepackage[normalem]{ulem}
\usepackage{xcolor}

\bibpunct{(}{)}{;}{a}{}{,}             
\usepackage{hyperref}
\hypersetup{
    colorlinks=true,
    linkcolor=blue,
    filecolor=magenta,      
    urlcolor=cyan,
    citecolor=blue,
    pdftitle={Overleaf Example},
    pdfpagemode=FullScreen,
    }

\begin{document}

   \title{The Impact of Axion-Like Particles on Late Stellar Evolution}

   \subtitle{From Intermediate-Mass Stars to core-collapse Supernova Progenitors}

   \author{
        I. Dom\'\i nguez\inst{1}
          \and
        O. Straniero\inst{2,3}
         \and
        L. Piersanti\inst{2,7}
         \and
        M. Giannotti\inst{4}
        \and
        A. Mirizzi\inst{5,6}
          }

   \institute{Departamento de F\'\i sica Te\'orica y del Cosmos, Universidad de Granada, E-18071 Granada, Spain
         \and
             INAF - Osservatorio Astronomico d’Abruzzo, Via Mentore Maggini snc, 64100 Teramo, Italy
         \and
             INFN - Sezione di Roma,  Piazzale Aldo Moro 2, I-00185 Roma, Italy 
          \and
            Centro de Astropart{\'i}culas y F{\'i}sica de Altas Energ{\'i}as (CAPA), Universidad de Zaragoza, Zaragoza, 50009, Spain
          \and
              Dipartemento Interuniversitario di Fisica ``Michelangelo Merlin'', Via Amendola 173, I-70126 Bari, Italy
          \and 
             INFN - Sezione di Bari, Via Orabona 4, 70126 Bari, Italy
             \and
              INFN - Sezione di Perugia,  Via A. Pascoli, 06123 Perugia, Italy
             }

   \date{Received XXX; YYY}

 
  \abstract
   {Stars with masses ranging from 3 to 11 M$_\odot$ exhibit multiple evolutionary paths. 
   Less massive stars in this range conclude their evolution as carbon-oxygen (CO) white dwarfs. 
   However, those that achieve carbon ignition before the pressure by degenerate electron 
   halts the core contraction may either form massive CONe/ONe white dwarfs, or 
   undergo an electron-capture supernova, or photo-disintegrate neon and proceed with further thermonuclear burning, ultimately leading to the formation of a gravitationally unstable iron core.}
   {An evaluation of the impact of the energy loss caused by the production of axion-like-particles (ALPs)
    on evolution and final destiny of these stars is the main objective of this paper.}
   {We compute various sets of stellar models, all with solar initial composition,
    varying the strengths of the ALP coupling with photons and electrons.}
   {As a consequence of an ALP thermal production, the critical masses for off-center C and Ne ignitions are both shifted upward.
   When the current bounds for the ALP coupling strengths are assumed, the maximum mass for CO WD progenitors 
   is about 1.1 M$_\odot$ heavier than that obtained without the ALP energy loss, while the minimum mass for 
   a core collapse supernova (CCSN) progenitor is 0.7 M$_\odot$ higher.  }
 {Current constraints from observed Type II-P supernova light curves and pre-explosive luminosity 
 do not exclude an ALP production within the current bounds. 
 However, the maximum age of CCSN progenitors, as deduced from the star formation rate of the parent stellar 
 population, would require a smaller minimum mass. 
 This discrepancy can be explained by assuming a moderate extra mixing  (as due to core overshooting or rotational induced mixing) above the fully convective core that develops during the main sequence.}  

   \titlerunning{ALPs production in 3 to 11 M$_\odot$ stars}
   \authorrunning{I. Dom\'\i nguez rt al.}
   
   \maketitle
%

\section{Introduction}

In this work we revise models of single stars with mass in the range $3 < M/$M$_\odot < 11$ taking into account 
the effects of a possible thermal production of axions or, more in general, axion-like-particles (ALPs).
In a previous paper, we have discussed the impact of ALPs on the evolution of star with $M>11$ M$_\odot$ \citep{straniero2019}, 
while the effect of a possible production of ALPs in intermediate-mass stars ($3<M/$M$_\odot<7$) 
has been illustrated in a pioneer work by \cite{Dominguez:1999gg}. 
 
Calculating stellar models with masses 3-11 M$_\odot$ is considerably more challenging than for higher masses,
due to the possible ignition of carbon and neon in environments with high or moderate electron degeneracy
\citep{garciaberro1997,ritossa1999,doherty2015, woosley2015,straniero2019_nic,limongi2024}. 
Very different scenarios characterize the late evolution of these stars. 
Intermediate-mass stars, i.e., those that never achieve conditions for carbon burning, evolve through the Asymptotic Giant Branch (AGB) and. eventually, they become carbon-oxygen white dwarfs (COWDs).
This is the most common type of WDs.
When they are part of a close binary, these WDs can accrete mass from a companion star and this 
accretion can trigger a variety of explosive phenomena, including: cataclysm variables, Novae and 
type Ia supernovae (SNIa). Therefore, the  maximum mass of the COWD progenitors is a fundamental parameter in 
studies of population synthesis and galactic chemical evolution.
After \cite{beckeriben1979}, it is often referred as $M_{up}$, but in this paper we prefer to use \Mup.   
The reason is that beyond this threshold, stars form a sufficiently massive carbon-oxygen (CO) core, enabling carbon ignition and the consignment production of oxygen and neon.
A further contraction of the core may or may not allow the heating necessary to activate the photo-dissociation of Ne and the following chain of $\alpha$, proton and neutron captures, at the end of which a iron-rich core will form \citep{woosley2015,straniero2019_nic,limongi2024}. 
This more advanced stage of the evolution may take place only if the CO core mass is above a critical value, 
roughly corresponding to the Chandrasekhar mass limit. 
Indeed, in case of a lower CO core mass, the concurrent action of pressure by degenerate electrons and plasma-neutrino cooling prevent the occurrence of a Ne burning.
 In practice, it exists a further critical stellar mass (hereinafter \Mupstar) above which stellar evolution 
 proceeds up to the formation of a gravitationally unstable iron core \citep{woosley2015,limongi2024}. 
 The final destiny of these massive stars  is determined by the collapse of the core, possibly followed by a supernova. Accordingly, the remnant may be a neutron star or a black hole.
 Instead, stars with mass in the range \Mup-\Mupstar skip the Ne burning and enter the super-AGB phase. 
 Depending on the mass-loss rate, which is expected to be quite strong along the super-AGB, the final destiny of these stars may be either a massive WD or an electron-capture Supernova. 
 As for the less massive COWDs, accretion in binaries may also result in various explosive phenomena or, 
 in case of a pure oxygen-neon WD (ONeWD), an  accretion-induced collapse (AIC) may occur, 
 whose result is a WD to neutron star (NS) transition. Binary systems hosting  accreting NSs are commonly 
 observed as low-mass-X-ray binaries (LMXBs).

In this paper we discuss the influence of the possible production of ALPs in stellar interiors and the consequent energy drain, on the values of the minimum stellar masses for carbon and neon burnings. 
In addition, we also show how this deviation from the standard physics may affects the second dredge up and the critical mass dividing stars undergoing a complete C burning from stars in which, after the off-center C ignition, the thermonuclear flame does not propagate inward down to the center. 
A pure ONe core forms only in case of a complete C burning, while an hybrid core 
(an inner CO core surrounded by an ONe mantel) is left if the C burning does not extend down to the center.   

Axion-like particles (ALPs) emerge naturally in a wide variety of extensions to the Standard Model (SM) of particle physics \citep{Jaeckel:2010ni,Ringwald:2014vqa,DiLuzio:2020wdo,Agrawal:2021dbo,Giannotti:2022euq,Antel:2023hkf}. These hypothetical pseudoscalar bosons are typically associated with the spontaneous breaking of approximate global symmetries and are characterized by their weak couplings and light masses. 
The most studied and theoretically motivated ALP is the QCD axion \citep{Peccei:1977hh,Peccei:1977ur,Weinberg:1977ma,Wilczek:1977pj}, originally proposed to solve the strong CP problem in quantum chromodynamics. While generic ALPs are not required to address this issue, they share many of the QCD axion’s properties and appear naturally in a wide range of well-motivated ultraviolet (UV) completions of the Standard Model.
From a top-down perspective, string theory compactifications predict the existence of a rich landscape of such particles—the so-called axiverse—which includes the QCD axion and an entire spectrum of ultralight ALPs spanning many orders of magnitude in mass \citep{Arvanitaki:2009fg,Cicoli:2012sz,Cicoli:2023opf}. These ALPs are expected to couple feebly to photons, fermions, and gluons, leading to distinctive phenomenological signatures across a range of experimental frontiers.
From a bottom-up approach, ALPs are of considerable interest due to their potential role in solving several open problems in cosmology and astrophysics. They are compelling dark matter candidates, particularly in the form of non-thermally produced cold relics via the vacuum realignment mechanism \citep{Abbott:1982af,Dine:1982ah,Preskill:1982cy,Arias:2012az,Adams:2022pbo}. In addition, they may help explain a variety of longstanding anomalies in stellar evolution and other astrophysical observations \citep{Giannotti:2015kwo,Giannotti:2017hny,Galanti:2022chk}. These include excessive energy losses in stars, unexplained features in white dwarf cooling, and anomalous transparency of the Universe to high-energy photons.
In this context, stars represent powerful laboratories for probing ALP properties. Stellar interiors host hot, dense plasmas where ALPs can be efficiently produced via processes such as the Primakoff effect, photon coalescence, or electron bremsstrahlung, depending on their couplings \citep{Raffelt:1996wa,Raffelt:1999tx}. Once produced, ALPs escape the stellar interior essentially unimpeded, carrying away energy and thereby altering stellar lifetimes and evolutionary tracks. This sensitivity has led to some of the most stringent constraints on ALP couplings to photons and electrons, particularly from observations of horizontal branch stars, red giants, white dwarfs, and supernovae \citep[see][for recent review]{caputoraffelt2024,carenza2025}.
The study of axions and ALPs in stellar environments thus offers a unique and complementary window onto new physics beyond the Standard Model. The interplay between theoretical developments and increasingly precise astrophysical measurements continues to sharpen our understanding of these elusive particles and their role in the cosmos.

ALPs are expected to affect astrophysical observations of stars in different evolutionary phases.
A stringent constraint to the ALP-photon coupling have been obtained by \cite{ayala2014} \citep[see also][]{Straniero:2015nvc},
comparing $R$ parameters\footnote{The $R$ parameter is the ratio of the number of HB stars and the number of bright RGB stars.}, as 
measured in a sample of  Galactic Globular Clusters, with the corresponding theoretical predictions. 
For the strength of the ALP-photon coupling, they found $g_{a\gamma} < 6.5\times 10^{-11}$ GeV$^{-1}$ at 95$\%$ CL.
A slightly more stringent bound, $g_{a\gamma} < 5.8\times 10^{-11}$ GeV$^{-1}$, has been obtained by the CAST collaboration \citep{CAST2021},  searching for solar axions. 
More recently, \cite{Dolan:2022kul}, analyzing a limited number of clusters for which $R_2$ measurements are also available, reported an even more stringent bound for the axion-photon coupling, i.e., $g_{a\gamma} \lesssim 0.47 \times 10^{-10}$~GeV$^{-1}$\footnote{The $R_2$ parameter is the ratio of the number of AGB stars and the number of HB stars.}.
However, due to the rapid acceleration of stellar evolution occurring after the central-helium exhaustion, the AGB phase is substantially short-lived. This results in a poor population of stars in this phase, which implies significant statistical fluctuations for $R_2$.
Constraints to the strength of the ALP-electron coupling $g_{ae}$ have been obtained from the luminosity of the RGB tip of Globular Clusters \citep{Viaux2013,capozzi2020,straniero2020}, from the period shift of pulsating WDs \citep{isern1992,corsico2012,kepler2021} and from the WD luminosity function \citep{millerbertolami2014,isern2018}.  
From the luminosity of the RGB tip, in particular, we get $g_{ae} < 1.5\times 10^{-13}$ \citep{straniero2020} and, by means of improved cluster distances \citep{2021MNRAS.505.5957B}, $g_{ae}<0.95\times 10^{-13}$ \citep{carenza2025}. 
Eventually, a new analysis based on Gaia-DR3 photometry and astrometry \citep{troitsky2025} reported $g_{ae}<0.52\times 10^{-13}$ (95\% CL). However, a detailed analysis of the overall error is missing from this work.
In the present study, we will assume values of the coupling strengths 
close to or smaller than the current bounds.

The aim of the present work is not the revision of these limits, but an analysis of the impact 
they have on the evolution of intermediate to massive stars.  
In Sec. 2 we describe the main features of our stellar evolution code and its setup, 
while in Sec. 3 the computational results are illustrated. In particular we will derive values of the critical masses, i.e., the minimum mass for the second dredge up (\msdu), the maximum mass for CO WD progenitors (\Mup) and the minimum mass for CCSN progenitors (\Mupstar). 
A final discussion follows in Sec. 4.

\section{Stellar evolution code and its setup}\label{sect2}
All the models presented in this paper have been computed by means of the Full Network Stellar Evolution code (FuNS). The FuNS version here adopted is the one described in Section 2 of \cite{straniero2019}. 
Due to their relevance for the interpretation of the values of the critical parameters determining the final destiny of stars, we provide here a brief description of the main code settings on which our calculations are based.
First, boundaries of the convective zones are fixed by means of the Ledoux criterion. Hence, no overshoot is applied to the external birder of the convective core that develops during the H-burning phase and no undershoot is considered at the inner border of the convective envelope during the red giant and asymptotic giant evolutionary phases.
Concerning the He-burning phase, induced overshoot and semiconvection are applied according to the procedure described in \cite{castellani1985}. Moreover, sporadic breathing pulses are blocked as in \cite{straniero2003}. Nuclear reaction rates are from the STARLIB repository \citep{starlib}. In particular, the triple-$\alpha$ and the $^{12}$C$+^{12}$C reaction rates are based on \cite{CF88} prescriptions, while the $^{12}$C$(\alpha,\gamma)^{16}$O reaction rate is based on \cite{deboer2017}. Eventually, effects of rotation are neglected.  
Noteworthy, for a given initial mass, this code setup 
results in a smaller CO core mass than obtained when an extramixing is applied 
at the external border of the convective core of an H burning star, as due to a 
convective overshooting or rotational induced mixing. On the other hand,
due to the adopted treatment of He-burning semiconvection and the $^{12}$C$(\alpha,\gamma)^{16}$O reaction rate, a reduced C/O ratio ($\sim 0.3$) is left in the core after the central-He burning phase \citep[see the discussion in][]{straniero2003}.

In the present work, we aim to evaluate the impact of
 a possible ALPs production on the evolution of stars developing a degenerate CO core
 after the He-burning phase. Therefore, energy-loss rates were computed following the prescriptions reported in the appendix of \cite{straniero2019}. ALP production via Primakoff effect, associated with photon-axion coupling $g_{a\gamma}$, and Compton, Bremsstrahlung, and $e^-e^+$-annihilation, due to  axion-electron coupling $g_{ae}$, were considered. 
 ALPs are assumed  massless, a valid approximation for $m_a<10$ keV.
These energy-loss rates depend on the strengths of these interactions. In the following, we also make use of two dimensionless parameters: 
$g_{10} \equiv g_{a\gamma}/(10^{-10}$ GeV$^{-1}$) and $g_{13} \equiv g_{ae}/ 10^{-13}$.
Current experimental and astrophysical constraints provides upper bound 
for these two parameters \citep[see the introdction and][ for more detailed reviews]{caputoraffelt2024,carenza2025}. 
To be conservative, we consider bounds obtained from studies based on a detailed error budget analysis, in practice, $g_{10}\leq0.6$ and $g_{13}\leq1.5$, for the axion-photon and the axion electron couplings, respectively.   
 
Thus, a large set of stellar evolutionary sequences was computed by adopting a solar chemical composition (\cite{lodders2009}), with  $Z_\odot=0.014$ and $Y_\odot=0.27$. 
Mass was varied between 3 and 11 $M_\odot$, in steps of 0.1 M$_\odot$.
Various couplings of ALPs with photons and electrons were assumed, namely: $g_{10}$ = 0.0, 0.2, 0.4, 0.6 and $g_{13}$= 0.0, 1.5. Some additional models were also computed assuming a higher axion-electron coupling strength, i.e., $g_{13}=4$.
All the evolutionary sequences starts from the pre-MS. For $M<$\Mupstar the calculations
was stopped at the onset of the first thermal pulse (AGB or super-AGB). For the more massive models, calculations were brought till the shell-Si Burning.

\section{Results}\label{sec_results}
\subsection{Stars that experience the second dredge up, skip the C burning and leave a CO WD}\label{sec_intermediatemass}

According to extant stellar models, only stars that develop a CO core with mass exceeding a critical value, \MCOcore$\sim 1.06$ M$_\odot$, attain the conditions for the C burning. In this context, the second dredge up (SDU) plays a fundamental role for the final destiny of intermediate mass stars.
When the central He is fully consumed, the CO core contracts and heats up, while its mass increases due to the helium burning in the overlying shell.
Meanwhile, a deep convective envelope develops, whose lower boundary is limited by the presence of an active H-burning shell. In principle, the mass of the CO core may grows until the He-burning shell gets close to the base of the H-rich envelope. However, the growing energy flux from the He-burning shell induces an expansion of the overlying layers, so-that the H-burning shell cools down and, eventually, dies down, allowing the penetration of the external convection. 
This episode limits the growth of the CO core and marks the end of the early-AGB phase \cite{beckeriben1979}. As a result, the SDU  prevents the carbon ignition in the core of intermediate mass stars, $3 < M/M_\odot < 7$. For instance, before the SDU,  
the He-rich core of a 6 M$_\odot$ model is $\sim1.31$ M$_\odot$. If the He-burning shell were free to advance up to the edge of this core, the final mass of the CO core would be much greater than the critical mass for carbon ignition. However, the SDU reduces the core mass down to $\sim 0.91$ M$_\odot$, thus preventing the carbon burning.
In practice, the C burning may only occur when the He-burning shell attains the critical CO-core mass, \MCOcore$\sim 1.06$ M$_\odot$,  before the occurrence of the SDU, as it happens in more massive stars. 
Intermediate mass stars are of pivotal importance for the galactic chemical evolution. After the  SDU, they enter the thermal-pulse AGB phase, during which the envelope is enriched with products of internal nucleosynthesis.  Thus, the intense stellar winds from these stars provides a significant contribution to the chemical enrichment of the galactic gas. Their compact remnants, the CO WDs, are the major constituent of the baryon dark matter in the Milky Way. In addition, interacting binaries hosting CO WDs give rise to important explosive phenomena, among with Cataclysm Variables, Novae and type Ia Supernovae. For all these reasons, the determination of the mass range of these stars is a fundamental task of stellar astrophysics. Unfortunately, this mass range is rather uncertain. First of all, different assumptions about the efficiency of the convective-core overshoot or rotational-induced mixing during the H-burning phase lead to a variation of the relation between the initial mass and the resulting CO-core mass. 
Also the amount of C left by the He burning, which depends on the treatment of semiconvection and overshooting during the He-burning phase and on the rates of relevant nuclear reactions (the triple$-\alpha$ and the $^{12}$C$(\alpha,\gamma)^{16}$O, in particular) affects the conditions for the C ignition and, in turn, the value of the critical core mass.  
With the standard settings of the FuNS code illustrated in the previous section, we find a minimum stellar mass for the occurrence of the SDU $M_{SDU}=4.3$ M$_\odot$\footnote{For solar composition stars.}, corresponding to a CO core mass after the SDU of 0.82 M$_\odot$. 
Instead, the more massive model in our sample that undergoes
 the SDU and skips the C burning is 7.4 M$_\odot$, corresponding to a 
 CO-core mass of 1.03 M$_\odot$.
\begin{figure}
  \centering
  \includegraphics[width=0.45\textwidth]{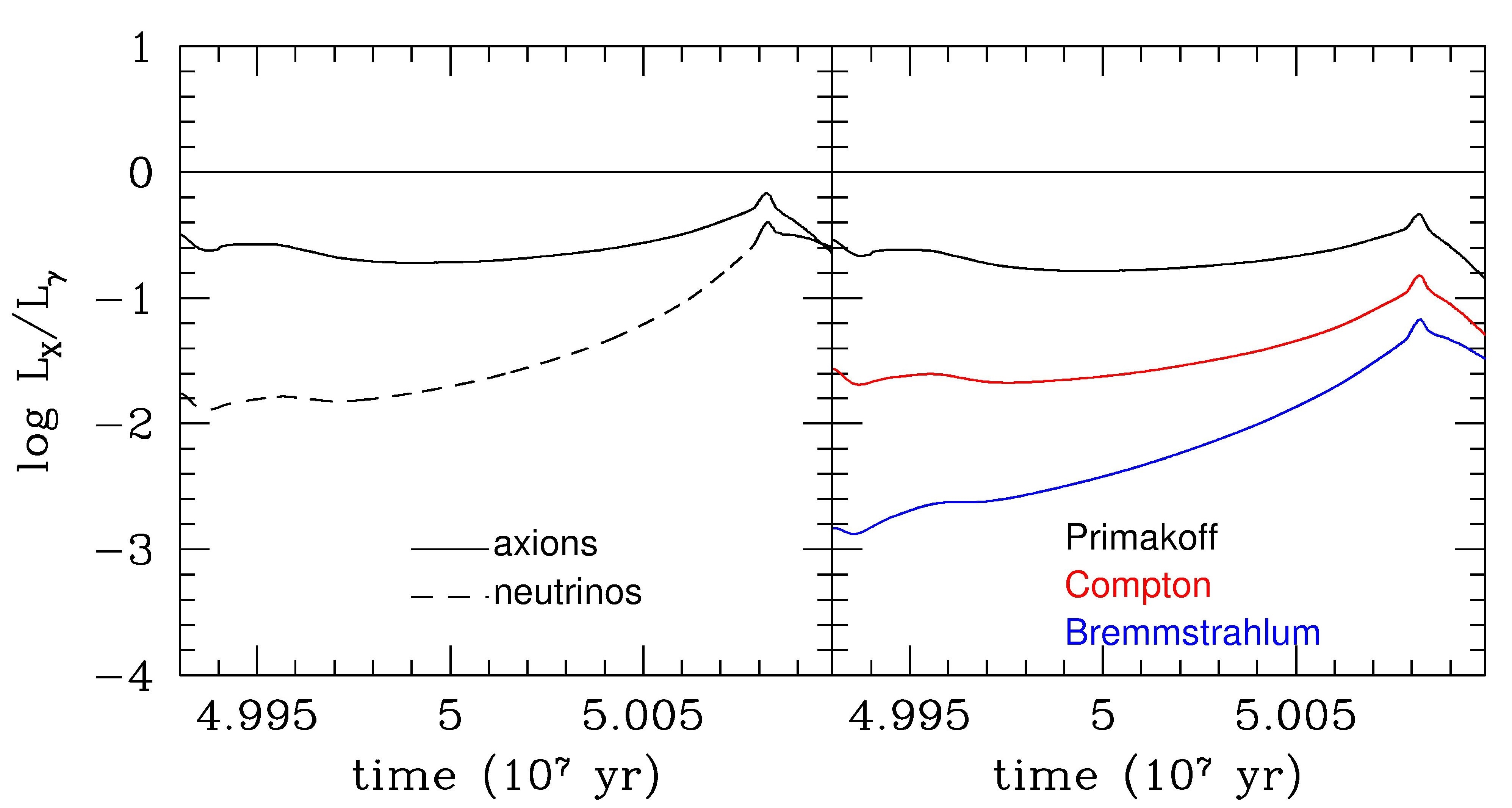}
   \caption{Left panel: evolution of the  ALP and neutrino luminosities through the late 
   portion of the He-burning phase and the subsequent early-AGB. Right panel: evolution of the contributions to the ALP luminosity of the different production processes. The plots refers to the 7 M$_\odot$ model with  $g_{10}=0.6$ and $g_{13}=1.5$.}
    \label{fig_aloss_7M}
\end{figure}

\begin{figure}
  \centering
  \includegraphics[width=0.45\textwidth]{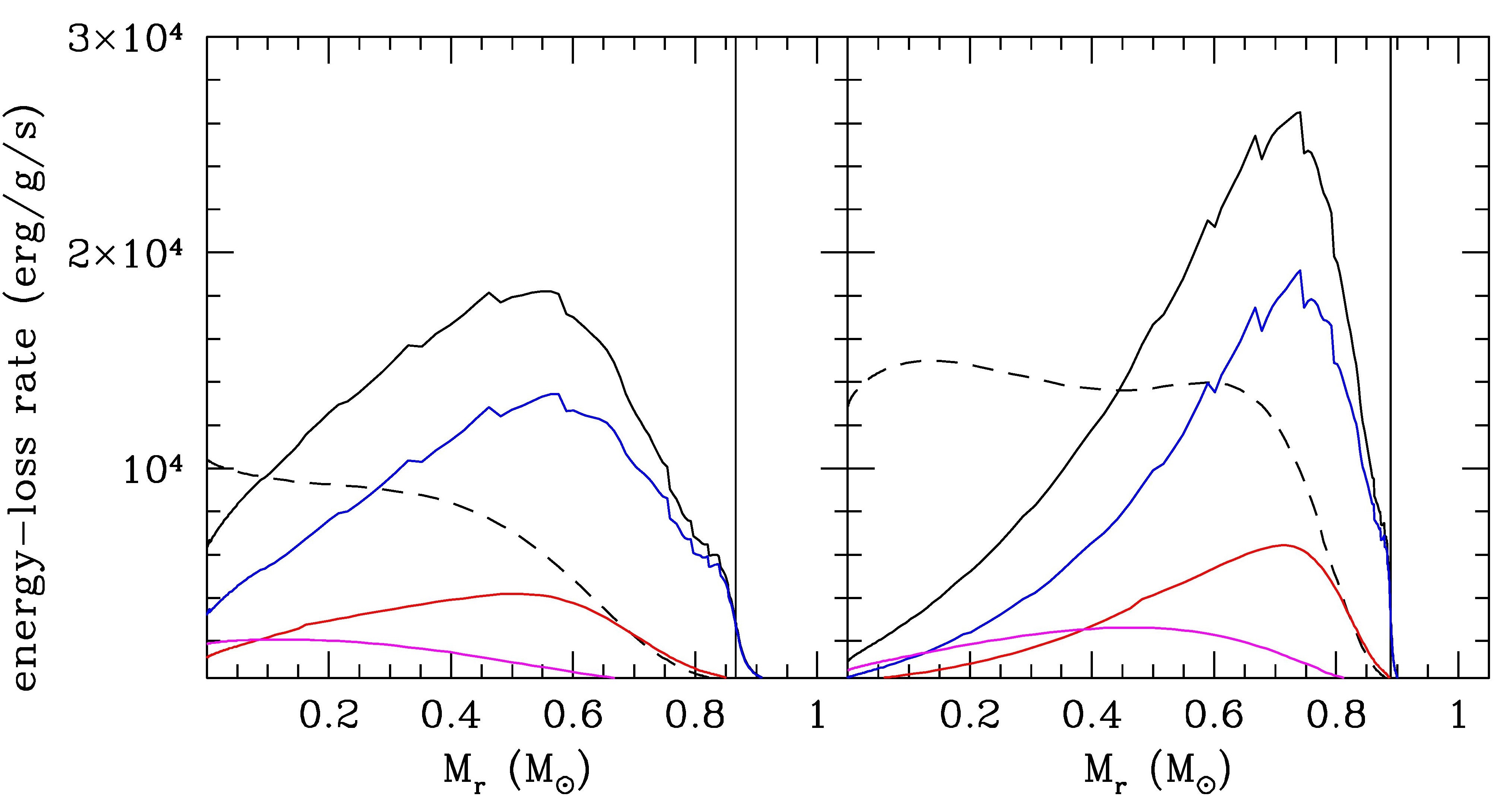}
   \caption{Energy-loss rates  due to ALPs production within the core
   of a 7 M$_\odot$ with  $g_{10}=0.6$ and $g_{13}=1.5$. The left and the right panels refer to the two blue points in Fig. \ref{fig_2dup_7M}, as taken before and after the SDU, respectively. The solid-black line is the total ALP energy-loss rate, while the blue, the red and the magenta lines represent the contributions of the Primakoff, Compton and Bremsstrahlung processes, respectively. For comparison, the neutrino energy-loss rate is also reported (black-dashed line).  
   }
    \label{fig_rates_7M}
\end{figure}

\begin{figure}
  \centering
  \includegraphics[width=0.45\textwidth]{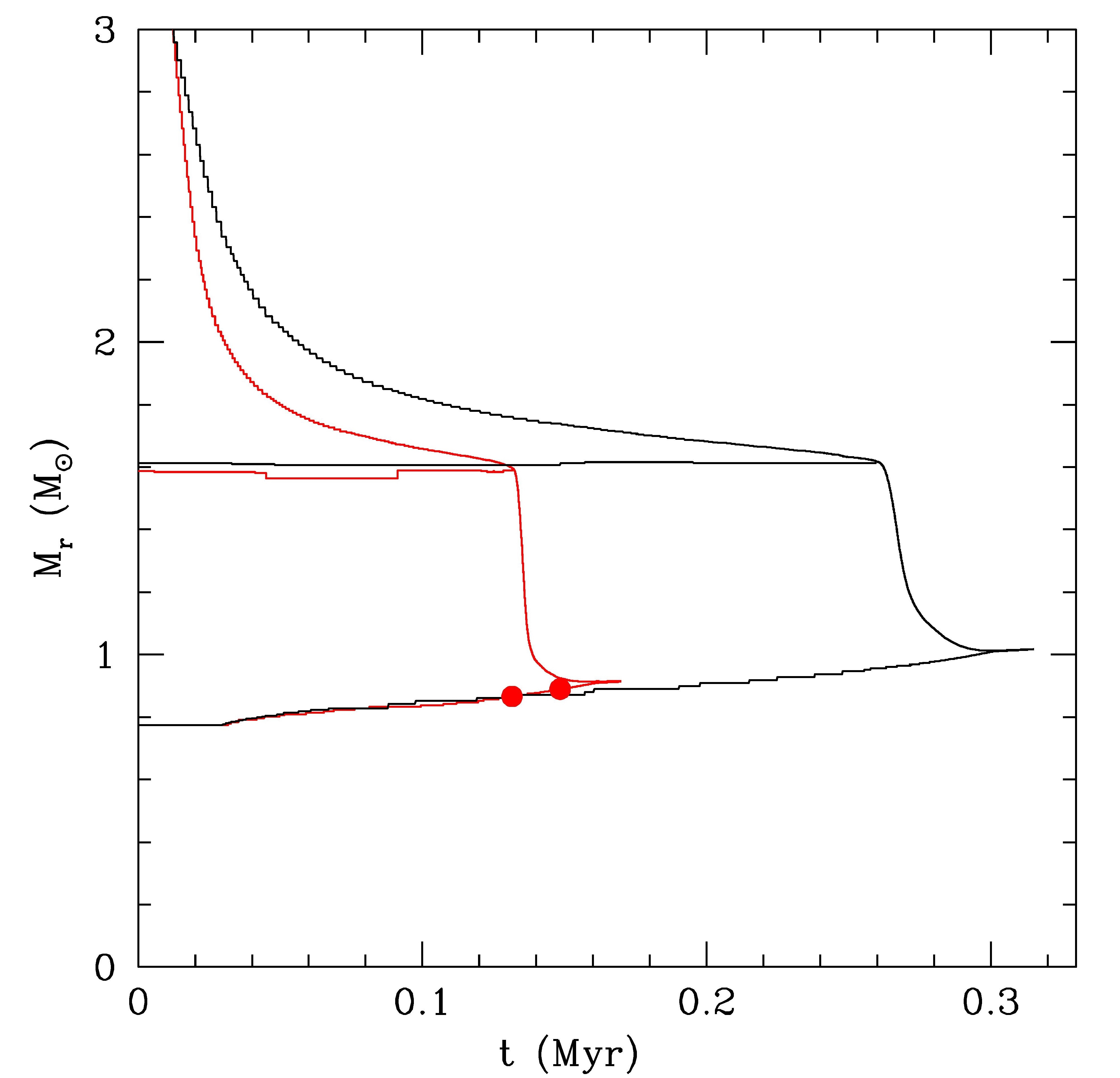}
   \caption{ALP impact on the SDU. 
   Black lines (top to botton): inner border of the convective envelope, 
   location of the H-burning shell and location of the He-burning shell, 
   for a 7 M$_\odot$ (no ALPs), during the early-AGB phase;
  red lines (top to bottom): same as the black lines, but with $g_{10}=0.6$ and $g_{13}=1.5$.
   The filled circles represent the He-shell locations for the two models in Fig. \ref{fig_rates_7M}
   }
    \label{fig_2dup_7M}
\end{figure}

 The inclusion of the energy loss possibly caused by a thermal production of ALPs affects both these mass limits. The energy-loss rates are steep functions of the temperature, so that the major influence of the ALPs production takes place during the He burning phase and beyond. Fig. \ref{fig_aloss_7M} shows the expected ALPS luminosity in a 7 M$_\odot$ model for $g_{10}=0.6$ and $g_{13}=1.5$. This choice of the couplings is representative of the current bounds. It results that the Primakoff process provides the major contribution to the energy loss. Noteworthy, the energy loss by ALPs is always greater than that by neutrinos. As a result, the He-burning lifetime is reduced and, in turn, the final He-core mass is smaller, while the carbon abundance in the CO core is slightly larger. This energy loss continues during the early-AGB. Fig. \ref{fig_rates_7M} reports the energy-loss rates of ALPs and neutrinos, in the core of the 7 M$_\odot$ model, before and after the SDU. As expected, the ALPs production is more efficient near the border of the CO core, where the temperature attains a maximum value. Therefore, as already noted by \cite{Dominguez:1999gg}, the shell-He burning proceeds at a faster rate and the stops of the H-burning, which coincides with the SDU, is anticipated (see Fig. \ref{fig_2dup_7M}). As a consequence, the minimum SDU mass is smaller than that obtained without ALPs, namely \msdu$=4.0$ M$_\odot$, which corresponds to a  a CO core mass after the SDU of 0.73 M$_\odot$. On the other hand, the more massive model that undergoes the SDU and skips the carbon burning has a larger mass, 8.5 M$_\odot$, and a larger CO core mass, 1.09 M$_\odot$.\\
 Noteworthy, the energy loss induced by an ALPs production reduces the stellar lifetime. In particular, a faster He consumption takes place   during the He-burning phase. As a result, the ALP inclusion allows the production of the more massive CO WDs in a shorter timescale, up to $\sim -25\%$  (see Tab. 1). 
 This occurrence have interesting implications, as, for example, a shorter delay time for the onset of the first type Ia supernovae.  

\subsection{Off-center carbon ignition in a degenerate CO core}\label{sec_cburning}
When the ALPs production is neglected, the CO core mass of models with $M\geq 7.5$ attains the critical value
 for the C burning before the occurrence of the SDU.
In the 7.5 M$_\odot$ model, an off-center C ignition takes place where the maximum in the temperature profile is attained, i.e.,  at $m_r=0.867$ M$_\odot$,  within a CO core of 1.064 M$_\odot$\footnote{We assume that the C burning starts when the local rate of nuclear energy release exceeds the energy-loss rate by neutrinos and ALPs.}. 
At the ignition point, the temperature and the density are 618 MK and $9.47\times 10^5$ g/cm$^3$, respectively. 
As noted since the pioneering work of \cite{beckeriben1979}, the critical CO core mass depends on several physics 
inputs, among which  the adopted nuclear reaction rates are the most relevant. 
Nonetheless, \cite{beckeriben1979} derived a critical CO core mass \MCOcore$=1.06$ M$_\odot$, practically the same as we find after more than 40 yrs. 

\begin{figure}
  \centering
  \includegraphics[width=0.45\textwidth]{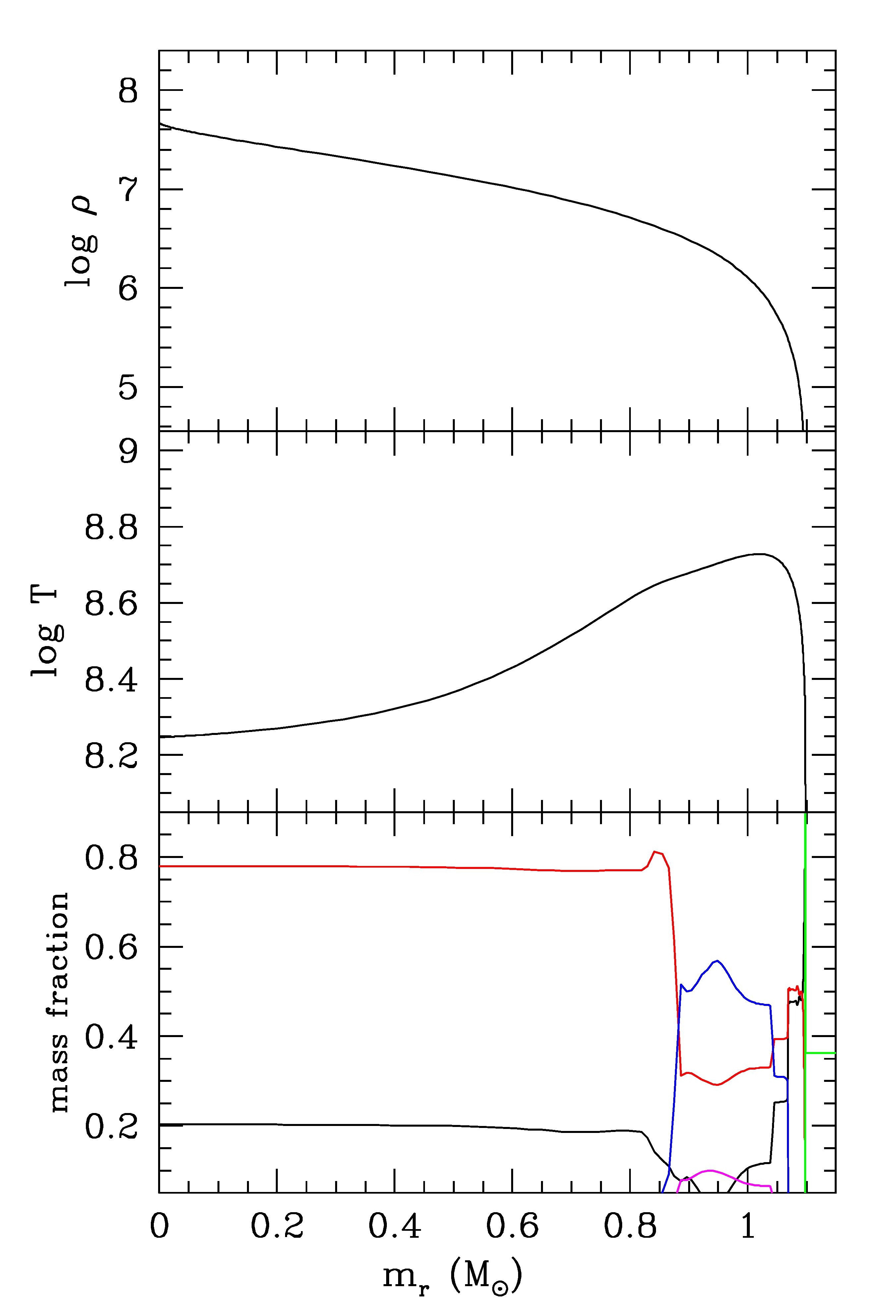}
   \caption{Top to bottom: density, temperature and chemical profiles within the core of a 7.7 M$_\odot$ model at the onset of the super-AGB phase. In the bottom panel, lines represent the mass fractions of $^{4}$He (green), $^{12}$C (black), $^{16}$O (red), $^{20}$Ne (blue) and $^{24}$Mg (magenta). No ALP production has been included in this model. Note that only above 0.85 M$_\odot$ the C burning has been active. This model represents an example of hybrid WD progenitors.}
    \label{fig_final_7p7}
\end{figure}

As the stellar mass increases, the C-ignition occurs  closer to the center. Then, after the off-center ignition, a convective shell develops.
Later on, the C burning moves inward through a series of progressively more internal C flashes. 
However, in models with mass between 7.5 and 7.7 M$_\odot$ the C burning extinguishes before reaching the center, thus leaving an unburned CO core surrounded by an ONe-rich mantel. Later on, these stars enters the super-AGB phase, during which they are expected to loose the H-rich envelope, thus leaving an hybrid WD, made of a CO core surrounded by a O-Ne mantel, whose mass ranges between 1.07 and 1.1 M$_\odot$.  
An example is reported in Fig. \ref{fig_final_7p7}, where the 3 panels show density, temperature and chemical composition profiles within the core of a 7.7 M$_\odot$ model entering the super-AGB phase.
Owing to  the sharp chemical gradient at the CO-ONe interface, a diffusive mixing of the core and the mantel material 
likely will occur. In case of accretion from a companion star in a binary system, these hybrid WDs may approach the Chandrasekhar-mass limit. Hence, the unburned carbon within the core may trigger a thermonuclear
explosion, rather than a core collapse. 

\begin{figure}
\centering      
\resizebox{0.45\textwidth}{!}{\includegraphics{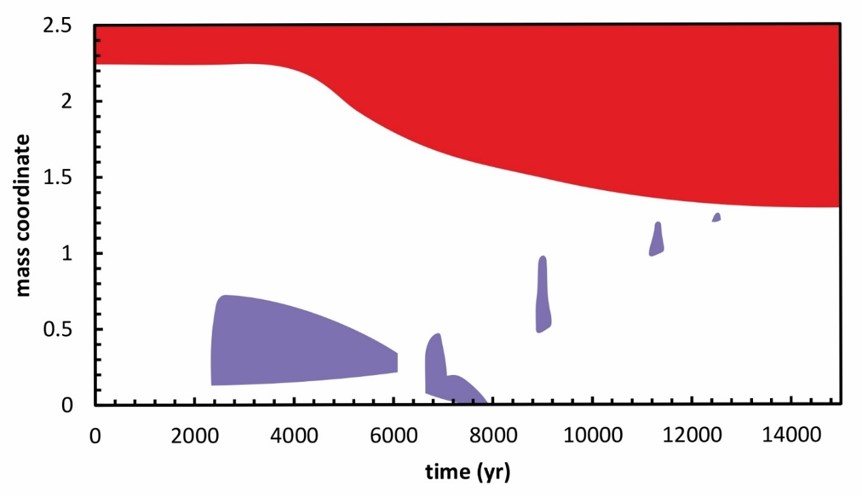}}
\caption{Kippenhahn diagram of the degenerate C burning in the 9 M$_\odot$ stellar model. The red region represents the convective envelope, while the violet regions are convective C-burning episodes. The t=0 point is arbitrary. Note that the SDU occurs after the first C flash.}
\label{fig_kip_9p0}   
\end{figure}

\begin{figure}
  \centering
  \includegraphics[width=0.45\textwidth]{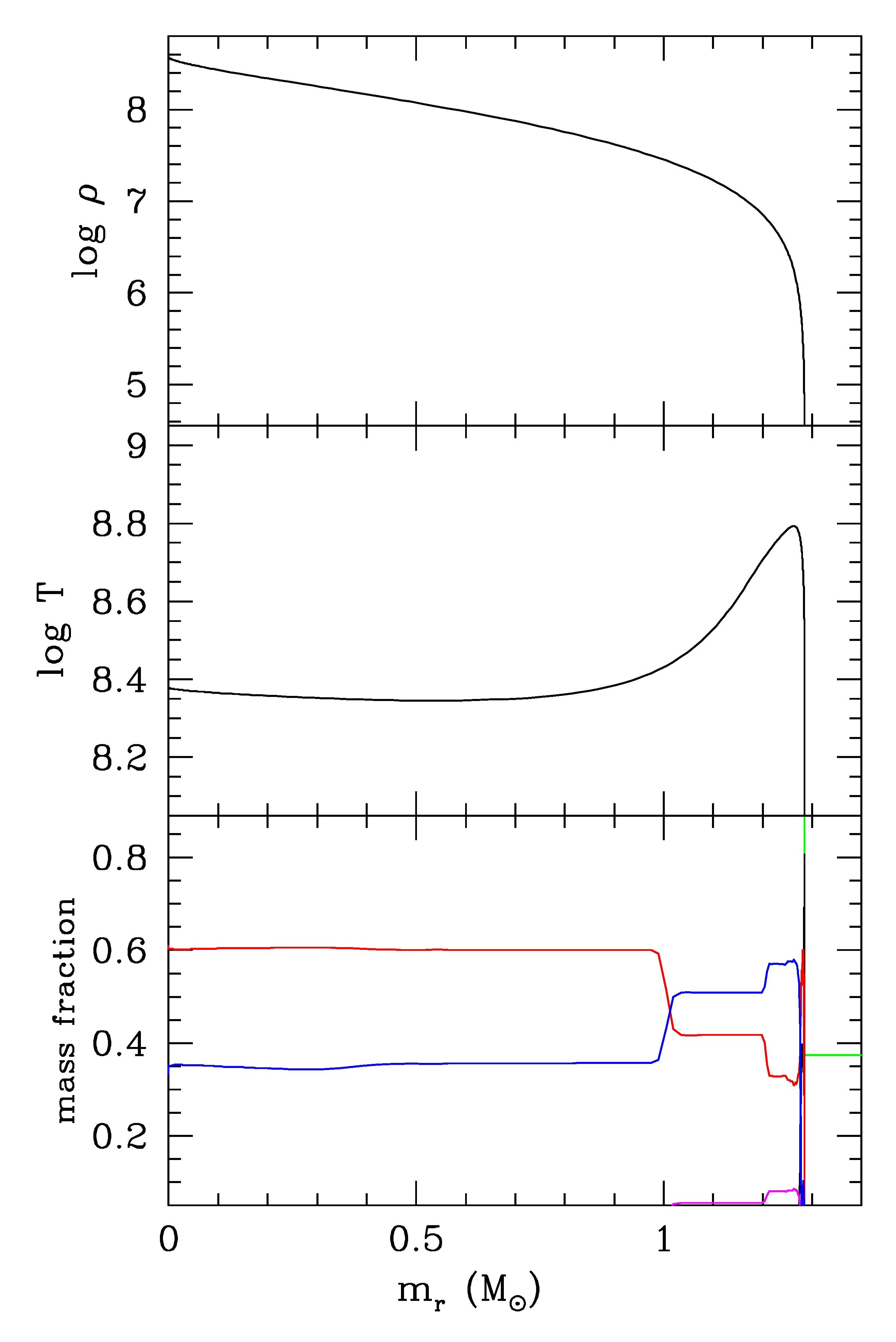}
   \caption{
   As in Fig. \ref{fig_final_7p7}, but for a 9 M$_\odot$ model (no ALPs). 
   }
    \label{fig_final_9p0}
\end{figure}

A complete C burning occurs in models with $M\geq 7.8$M$_\odot$.  
The development of the convective shells driven by the various C-flashes in the 9 M$_\odot$ model 
is illustrated in Fig. \ref{fig_kip_9p0}. 
The first C ignition occurs at $m_r\sim 0.15$ M$_\odot$. Suddenly a convective shell develops that extends up to  $m_r\sim 0.7$ M$_\odot$.
Then, as C is consumed, the convective shell shrinks and, eventually, disappears. The consequent contraction induces a second C flash, at $m_r\sim 0.11$ M$_\odot$, coupled to a new convective shell partially overlapping the previous one. 
Later on, the inner border of this convective shell moves slowly inward, until a central C burning sets in, followed by a series of three weaker and more external C flashes. The resulting core composition is shown in the lower panel of Fig. \ref{fig_final_9p0}.

\begin{figure}
  \centering
  \includegraphics[width=0.45\textwidth]{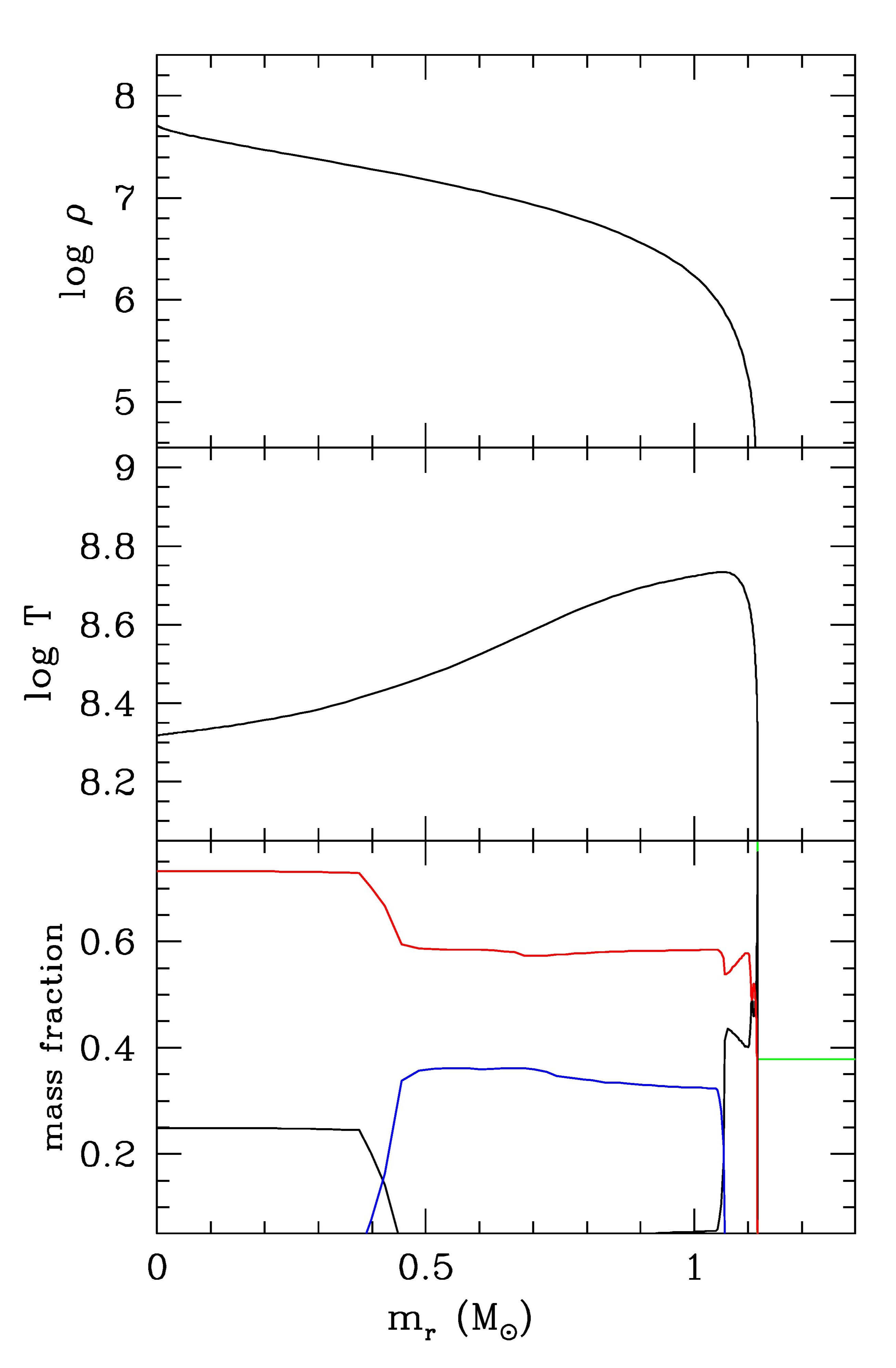}
   \caption{
   As in Fig. \ref{fig_final_7p7}, but for a 8.6 M$_\odot$ model ($g_{10} = 0.6$ and $g_{13} = 1.5$). 
   Only above 0.4 M$_\odot$ the C burning has been completed.
   }
    \label{fig_final_8p6ax}
\end{figure}

 As illustrated in the previous section, the ALPs production moves upward both \Mup and the corresponding \MCOcore. The values of the critical masses, as obtained under different assumptions about the strengths of the ALP couplings with photons and electrons,  are reported in Tab. 1. The total stellar lifetime and the central C mass fraction after the He-burning phase are also shown. 
When only the axion-photon coupling is considered, namely $g_{10} = 0.6$ and $g_{13} = 0$,   \Mup is 8.4 M$_\odot$ that corresponds to a CO-core mass  $1.095$ M$_\odot$. 
In this case, the C ignition occurs at a mass coordinate $m_r=0.66$ M$_\odot$, where $T=630$ MK and $\rho=3.02\times 10^6$ $g/cm^{-3}$, both higher than the values obtained without ALPs. 
When also the coupling with electrons is switched on, namely $g_{10} = 0.6$ and $g_{13} = 1.5$, the critical mass rises at 
\Mup$=8.6$ M$_\odot$, while \MCOcore$=1.104$ M$_\odot$. In this case, the C ignition occurs at $m_r=0.66$ M$_\odot$, where $T=632$ MK and $\rho=2.94\times 10^6$ $g/cm^{-3}$.

In addition to the increase of \Mup, the ALPs production also implies a higher minimum CO core mass for the C ignition (Tab.1). This is a direct consequence of the additional energy loss in the CO core of a star approaching the C ignition, only partially counterbalanced by the slightly higher C mass fraction left in the core after the He burning (see the last column  in Tab. 1.

As for the standard models, C burning does not extend down to the center for models with mass slightly above  \Mup. In particular, if   $g_{10} = 0.6$ and $g_{13} = 1.5$, an incomplete C burning occurs for masses between 8.6 and 8.8 M$_\odot$ (see Fig. \ref{fig_final_8p6ax}).

\begin{figure}
    \centering
   \includegraphics[scale=0.1,width=8cm]{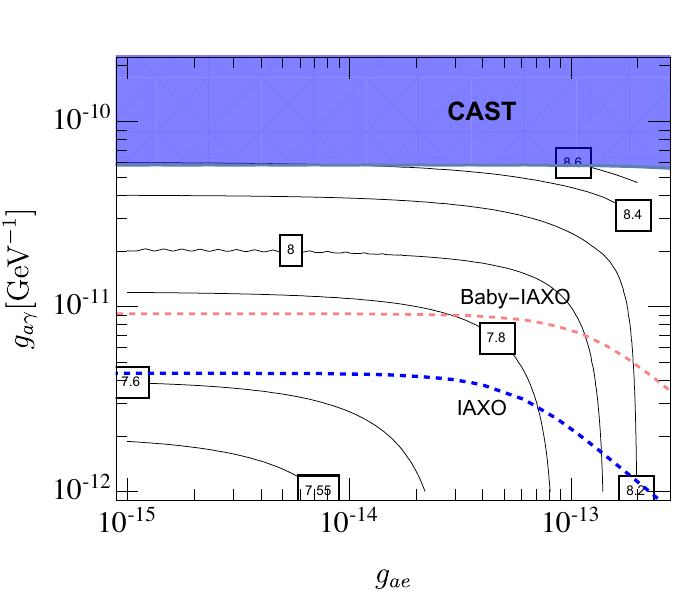}
\caption{Contours of $\Delta$ \Mup, in steps of 0.2 M$_\odot$ in the $g_{a e}$ - $g_{a\gamma}$ plane, considering the current bounds. 
		The projected sensitivity of Baby IAXO and IAXO are also indicated for reference, as it is the area excluded by the CAST result.}
\label{fig:Mup}
\end{figure}

In Fig.~\ref{fig:Mup}, the variation of \Mup is shown in the 2 parameters space, $(g_{13}.g_{10})$. The shaded-blue band represents the $g_{10}$ values excluded by the CAST helioscope measurements \citep{CAST2021}. For comparisons, the expected sensitivities of BabyIAXO and IAXO are also reported \citep{IAXO2024, IAXO2020, Armengaud:2014gea, Giannotti:2016drd, Armengaud:2019uso}. Practically, if the coupling of ALPs to electrons is negligible, the bound derived from CAST excludes that the axion-photon coupling could enhance \Mup above $\sim 8.4$ M$_\odot$. More stringent limits will be achieved by Baby-IAXO and, eventually, by IAXO.

\begin{table}
 \begin{center}
 \caption{Impact of ALPs on the minimum mass for off-center C ignition. The 6 columns list, respectively, the
 ALP-photon coupling strength, the ALP-electron coupling strength, \Mup, and the corresponding CO core mass, total lifetime and central C mass fraction after the core-He burning.}
\begin{tabular}{c c c c c c} 
\hline\hline
$g_{10}$  & $g_{13}$ & \Mup & $M_{CO}$ & Age & $X_C$  \\
\hline  
0.0  & 0.0 & 7.5 & 1.064 & 43.5 & 0.203\\  
0.0  & 1.5& 8.0&  1.089 & 37.9  & 0.216\\ 
\hline
0.2  & 0.0 & 8.0 & 1.101 & 37.9 & 0.216\\
0.2  & 1.5& 8.2 & 1.104 & 36.0 & 0.227\\
\hline
0.4  &  0.0 & 8.2 & 1.099 & 35.9 & 0.225\\
0.4  &  1.5& 8.4 & 1.104 & 34.1 & 0.241\\
\hline
0.6  & 0.0  & 8.4  & 1.095 & 33.9 & 0.235\\
0.6  & 1.5  & 8.5  & 1.104 & 33.1 & 0.241 \\
\hline 
0.6& 4.0& 9.2 & 1.155 & 28.2 & 0.239\\
\hline\hline
\end{tabular}
\tablefoot{Masses are in M$_\odot$, ages in Myr.}
\end{center}
\label{tab1}
\end{table}


\begin{table}
\begin{center}
\caption{Impact of ALPs on the minimum mass for CCSN progenitors. The 7 columns list, respectively, the
 ALP-photon coupling strength, the ALP-electron coupling strength, \Mupstar, the corresponding total lifetime, the final luminosity, Ne core mass and CO core mass.}
\begin{tabular}{c c c c c c c}
 $g_{10}$  &  $g_{13}$  & \Mupstar & Age & log L/L$_\odot$ & $M_{Ne}$ & $M_{CO}$ \\ 
\hline\hline
0.0  & 0.0 & 9.7 & 25.85 & 4.39 &  1.378  &  1.393\\   
0.0  & 1.5 & 9.9 & 24.84 & 4.41 & 1.375 &  1.387 \\  
\hline
0.6 & 0.0 & 10.3& 22.90  & 4.41 & 1.376 & 1.391  \\
0.6  & 1.5 & 10.4& 22.49 & 4.43 & 1.376 & 1.392 \\
\hline\hline
0.6  & 4.0 & 11.1 & 19.99 & 4.42 & 1.389 & 1.427  \\
\hline
\end{tabular}
\tablefoot{Masses are in M$_\odot$, ages in Myr.}
\end{center}
\label{tab2}
\end{table}

\subsection{Off-center Ne ignition and Core-Collapse SNe progenitors}

As with the C burning, it exists a minimum core mass above which the temperature increases enough to switch on the Ne photodisintegration through the $^{20}$Ne$(\gamma,\alpha)^{16}$O reaction.
Stars that undergoes a complete C burning, at the end of which the ONe core mass is lower than this threshold, enters the super-AGB phase, during which an intense super-wind erodes the H-rich 
envelope. In principle the core mass can grow during this evolutionary phase and eventually 
it might approach the Chandrasekhar limit. In this case, a core collapse would occur followed by a supernova (electron-capture SNe). However, many of these stars likely leave the super-AGB before this condition is attained, ending up as an ONe WD with mass between 1.15 and 1.35 M$_\odot$.  
In close binary, these WDs can accrete mass,
giving rise to Nova-like phenomena or even  experience an accretion induced collapse (AIC).

More massive stars develop core masses above the threshold of Ne 
photodisintegration and, according to the extant models, they are expected to complete all nuclear burnings up to the formation 
of an iron-rich core core \citep[e.g.,][]{woosley2015, straniero2019, limongi2024}).
Then, the iron core will collapse and a classical CCSNe may occur. 

\begin{figure}
  \centering
  \includegraphics[width=0.45\textwidth]{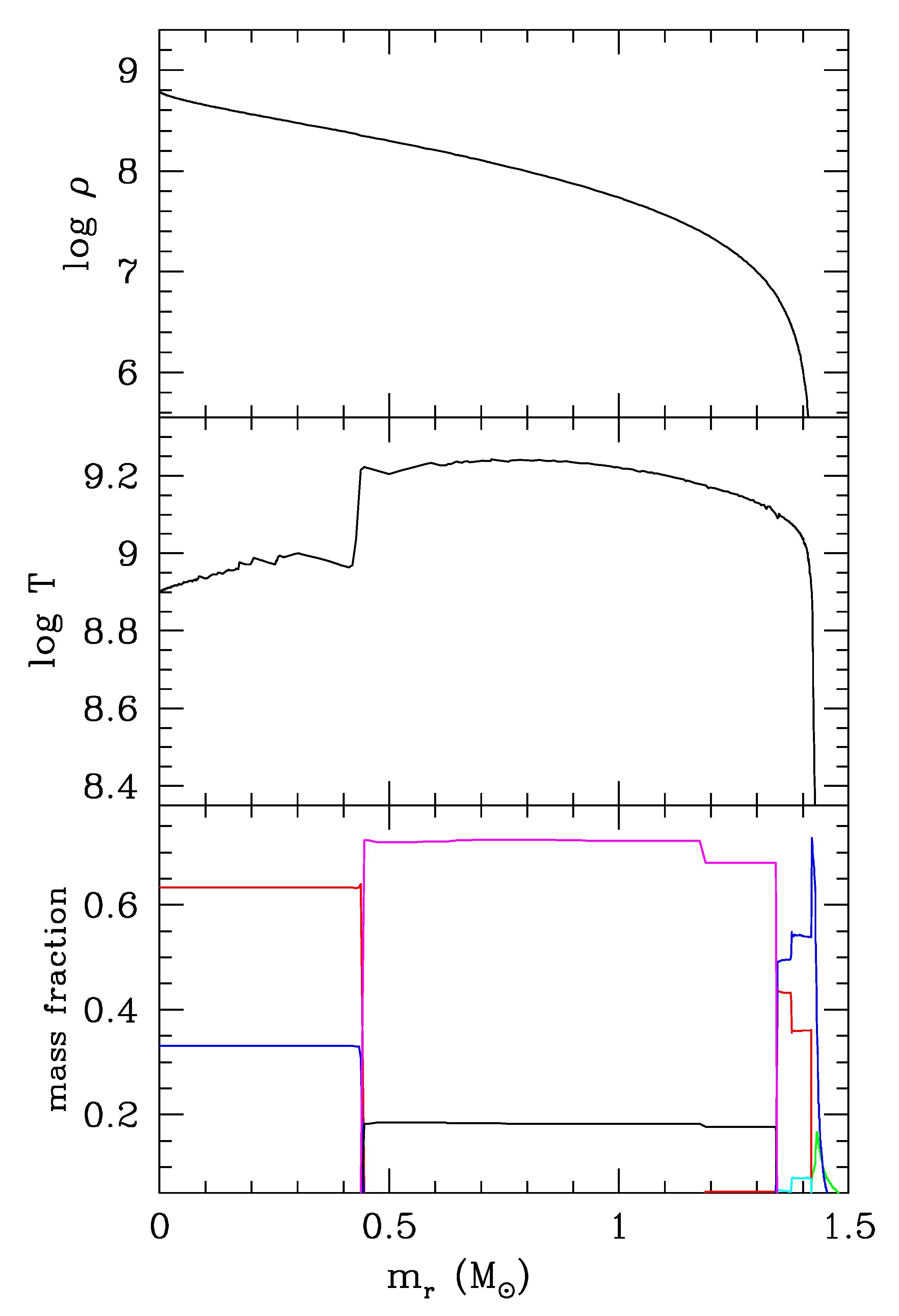}
   \caption{Top to bottom: density, temperature and chemical profiles within 
   the core of a 10 M$_\odot$ model during the advanced burning phase (Ne burning 
   and beyond). 
   In the bottom panel, lines represent the mass fractions of $^{16}$O (red), $^{20}$Ne (blue), $^{24}$Mg (cyan), $^{28}$Si (black) and $^{32}$S (magenta).
    No ALP production has been included in this model. 
    The burning front is located at $m_r\sim 0.44$ M$_\odot$. Above this point and 
    up to $m_r\sim 1.33$ M$_\odot$, oxygen and neon have been already fully consumed 
    and the main constituents are now silicon and sulfur. 
The penetrating burning front is at $m_r\sim 0.44$ M$_\odot$. More outside, oxygen and neon are already fully consumed and the main constituents are silicon and sulfur.}
    \label{fig_final_10p0}
\end{figure}

When the ALP cooling is neglected, we found 
that the minimum stellar mass igniting Ne is 
\Mupstar$=9.7$ M$_\odot$. In this case, the Ne-ignition occurs at a mass coordinate 
of 0.86 M$_\odot$, within an ONe core of 1.378 M$_\odot$ and a CO core of 1.393 M$_\odot$. 
The off-center Ne burning begins with the endothermic $^{20}$Ne$(\gamma,\alpha)^{16}$O reaction  
and proceeds with a chain of $\alpha$ and proton captures, while the inverse processes become increasingly efficient as the temperature rises.
Meanwhile, the burning flame moves slowly inward.
For comparisons, \cite{limongi2024} found \Mupstar$=9.22$ M$_\odot$. 
Their $M=9.3$ M$_\odot$ model, in particular, is the most similar to our 9.7 M$_\odot$.   
At the Ne ignition, the ONe and the CO cores of this \cite{limongi2024} model are, respectively, 1.367 and 1.382 M$_\odot$.
The difference in the minimum Ne burning mass may be likely ascribed 
to the inclusion in \cite{limongi2024} of a core overshoot during the H-burning phase and to the algorithm they use to damp breathing pulses appearing during the late part of the He-burning phase. In practice, at the onset of the C burning, the central C mass fraction is $\sim 0.4$ in the \cite{limongi2024} model, about double what we find. Fig. \ref{fig_final_10p0} shows the density, temperature and chemical profiles, within the core of a 10 M$_\odot$ model during the advanced burning phases, when the thermonuclear flame triggered by the Ne photodisintegration moves inward.  

Next, we computed additional models, incorporating ALP energy loss.  
We found that, for  $g_{10}$ = 0.6 and $g_{13}$ = 1.5, \Mupstar increases to 10.4 M$_\odot$. In this model, Ne firstly ignites at a mass coordinate of 
0.858 M$_\odot$, within an ONe core of 1.381 M$_\odot$ and a CO core of 1.408 M$_\odot$.  
Results for different choices of the coupling strengths are reported in Tab. 2. 

With or without ALPs, Ne-ignition always occurs when the CO core mass  exceeds 1.39 M$_\odot$ (or ONe core mass $>1.37$ M$_\odot$). Instead the corresponding initial masses moderately depends  on the axion couplings. The minimum mass for CCSNe progenitors increases by maximum 0.8 M$_\odot$, about  8\%, when ALP coupling strengths within current bounds are assumed. 

\subsection{Age, luminosity and mass of the lightest CCSN progenitor}

Numerous approaches have been developed to estimate the mass of CCSN progenitors.
A division into two classes can be made: one comprises methods based on 
explosive outcomes (light curves and spectra), 
while the other includes those relying on observed progenitor properties 
(pre-explosive luminosities or the ages of the parent stellar population).
In both cases, theoretical models — explosive in the first case and hydrostatic 
in the latter — must be used to relate the observed properties to the progenitor
initial mass (or ZAMS mass). This makes the result highly dependent on the model used.
Keeping this warning in mind, we will now compare the results of these studies 
to our theoretical prediction.
 
As shown by \cite{barker2023}, the progenitor mass of a SNII-P 
can be estimated from the observed bolometric light curves. In particular, they found 
a relationship between the plateau luminosity and the progenitor iron-core mass.
Hence, using a theoretical initial mass-final core mass relation, \cite{barker2023}
derived a minimum ZAMS mass of $9.8^{+0.37}_{-0.27}$ M$_\odot$ for SNII-P progenitors, 
which is in excellent agreement with the value we derived for the reference model, 
and does not exclude a thermal production of ALPs with electron and photon coupling 
within the current bounds (see Tab. 2).

A method based on the  luminosity of red-super-giant (RSG) progenitors 
has been developed by \cite{sm2009,sm2015}. 
Based on 20 archival pre-explosive images, they found
a minimum mass of CCSN progenitors of $9.5^{+0.15}_{-1.0}$ M$_\odot$, 
also in agreement with the prediction of our reference model. However, this result 
seems to exclude a sizable additional cooling due to a possible thermal production 
of ALPs.
In this case, the observed parameter is the RSG luminosity, while the 
progenitor mass is obtained from a specific set of stellar evolutionary tracks. 
If the ALP production is effective in stellar interiors, the progenitor mass 
would need to be derived using a set of models that incorporates the associated 
energy drain. This inconsistency can be resolved by directly comparing 
observed RSG luminosities with the corresponding predicted values.
The dimmer RSG progenitors in the CCSN sample,  
SN 2003gd, 2005cs, 2008bk, 2009md and  2012A (see  ~\cite{sm2015}), 
have luminosities ($\log L/L_\odot$) ranging from 4.4 to 4.6. 
These observed minimum luminosities are well  reproduce by our models, with or without ALPs.  
Note that, while the minimum progenitor mass increases when ALPs are produced 
in stellar interior,
the corresponding final luminosity is marginally affected. Indeed, as explained in \cite{straniero2019}, models with ALPs show a fainter progenitor. For this reason, the larger minimum mass obtained when ALP energy loss is switched on is 
compensated by a reduction of the  final luminosity.   

A last method employed to estimate the minimum mass of supernova progenitor 
is based on the age of the parent stellar population. Recently, \cite{diaz_2021}
investigated the local star formation histories to estimate the ages of 
a sample of 22 historic CCSNe. In this way, they inferred the slope of the age 
distribution and the maximum age of CCSNe. Hence, by 
using single-star evolutionary tracks, they transform the progenitor age 
distribution into a progenitor mass distribution. 
The resulting minimum mass for CCSNe is $8.60^{+0.37}_{-0.41}$. 
This value is smaller than limits established by other methods, 
as well as our theoretical predictions. Note that the adopted stellar models 
\citep{marigo_2017} have been computed assuming a large convective-core overshoot for H-burning stars that implies a lower ZAMS mass for a fixed stellar lifetime. To avoid the dependence on the adopted stellar models, we can directly compare the age inferred from the star formation history  with the one we predict for the minimum CCSN progenitor. In particular \cite{diaz_2021} obtained a maximum age $35.1^{+4.1}_{-3.0}$ Myr, to be compared with the 25.85 Myr  of our 9.7 M$_\odot$ reference models (see Tab. 2). The discrepancy is even worst when a thermal production of ALPs is considered. Indeed, AlP energy loss accelerates fuel consumption during the main nuclear burning phases.

\section{Summary and Conclusions}
In the present study we have investigated the impact of ALPs  
on the final destiny of stars with masses ranging between 3 and 11 M$_\odot$.
In particular, we have derived the minimum stellar mass for the occurrence of the second dredge up
and the critical masses for off-center carbon and neon ignitions in degenerate cores, under various assumptions
about the strengths of the ALP-photon and ALP-electron couplings.  
The final destiny of stars in this mass range is of pivotal importance in astrophysics. Below the mass 
threshold for Ne ignition, these stars leave massive WDs that, in binary systems,  may be accreted via mass
 transfer from the companion star, giving rise to classical or recurrent Novae, 
 prompt SNIa and AIC events. Above this threshold,
they are progenitor of CCSN SNe.  

Our main findings are summarized below, highlighting the differences between reference models 
(without ALP production) and models calculated with ALP production rates at the current upper 
coupling strength limits, i.e., $g_{10} = 0.6$ and $g_{13} = 1.5$.   
In all cases, the initial chemical composition is Y=0.27 and Z=0.014, while both convective overshoot and rotation are neglected.

\begin{enumerate}
\item \Mup, the minimum initial stellar mass that experiences C-ignition is 
7.5 M$_\odot$ and 8.5 M$_\odot$, for the reference and the ALP models, respectively. 
The increase of \Mup is due to the earlier occurrence of the second dredge up in ALP models, 
which is a consequence of extra energy loss within the He-shell. 
The maximum mass of  the corresponding CO WDs increases  by 0.04 M$_\odot$, from 1.06 to 1.10 M$_\odot$, 
while the lifetime of the corresponding WD stellar progenitors is reduced from above 43.5 to 33.1 Myr. 

\item The lifetime of the more massive CO WD ($0.8 < M_{WD}/$M$_\odot < 1.1$) is shorter when ALPs are included
 in the calculation of stellar models. The lifetime reduction is more pronounced at smaller masses. For instance, for a 0.88 M$_\odot$ CO WD progenitor, it is reduced from 113 to 50 Myr.  This occurrence might have an impact on the time elapsed between the progenitor formation and the explosion of the first type Ia supernovae which is expected to be  $\geq 40$ Myr \citep{rodney_2014}. 

\item Models whose mass is slightly above \Mup undergo an incomplete C burning. After the off-center ignition,
the burning flame moves inward, but never reach the center. These models leave an hybrid core,
 whose innermost portion is made of C and O, while the outermost is a mixture of O and Ne.  
 The minimum mass in which the C burning is completed (up to the center) is 8  and 8.9  M$_\odot$, for models
 without and with ALPs, respectively. In both cases, the corresponding core mass is $\sim 1.15$ M$_\odot$. 
 If the core mass of these stars approaches the Chandrasekhar limit, due to shell burning in the super-AGB phase, 
 or the mass of the resulting WD increases, due to accretion in binary systems,
 the presence of carbon near the center would likely trigger a thermonuclear explosion rather than a core collapse.
 
\item More massive models form an O-Ne core surrounded by an active C burning shell. Later on, those with mass above
\Mupstar ignite Ne. This further threshold is 9.7 M$_\odot$, for no-ALP models, and up to 10.4 M$_\odot$ when ALP energy loss is included. Stars forming an ONe core whose mass is lower than \Mupstar will form ONe WDs or, in case
the core mass grows up to the Chandrasekhar limit during the super AGB, their outcome will be an electron-capture SNe. Then, accretion in binary may trigger an AIC. In any case, the mass range of these stars is shifted upward when ALPs are considered. Therefore, if a Salpeter-like mass distribution is assumed, i.e. a power law with exponent $\alpha=-2.35$, the number of these stars is reduced by $\sim 30\%$ when ALPs are  included in stellar model calculations.

\item \Mupstar is also the minimum mass for progenitors of CCSNe. Both the observed minimum luminosity 
of CCSN progenitors \citep{sm2015} and the mass estimated from the plateau luminosity of SNII-P \citep{barker2023} 
are compatible with the predictions of our models, either with or without ALPs. On the contrary, 
the maximum age of the parent stellar population \citep{diaz_2021} appears definitely higher than that required by our models.
This occurrence might imply the need of a convective core overshoot or rotational 
induced mixing during the main sequence phase. Noteworthy, the lifetime of \Mupstar model 
is substantially shorter in case of ALPs (see Tab. 2). Noteworthy, if asteroseismology can independently determine the extent of mixing in massive main-sequence stars, then the measured ages of lighter core-collapse supernova progenitors could be used to constrain the coupling of ALPs with standard particles. 
 \end{enumerate}  

Let us finally note that our ALP stellar models do not present the suppression of the blue loop during 
the core-He burning phase, as previously reported by \cite{friedland2013}.

\begin{acknowledgements}
 This article is based upon work from COST Action COSMIC WISPers CA21106, 
supported by COST (European Cooperation in Science and Technology); 
I.D acknowledges founding from the project PID2021-123110NB-I00 financed by the Spanish 
MCIN/AEI /10.13039/501100011033/ \& FEDER A way to make Europe, UE.
MG acknowledges support from the Spanish Agencia Estatal de Investigación under grant PID2019-108122GB-C31, funded by MCIN/AEI/10.13039/501100011033, and from the “European Union NextGenerationEU/PRTR” (Planes complementarios, Programa de Astrofísica y Física de Altas Energías). He also acknowledges support from grant PGC2022-126078NB-C21, “Aún más allá de los modelos estándar,” funded by MCIN/AEI/10.13039/501100011033 and “ERDF A way of making Europe.” Additionally, MG acknowledges funding from the European Union’s Horizon 2020 research and innovation programme under the European Research Council (ERC) grant agreement ERC-2017-AdG788781 (IAXO+). 
The work of A.M.  was partially supported by the research grant number 2022E2J4RK "PANTHEON: Perspectives in Astroparticle and
Neutrino THEory with Old and New messengers" under the program PRIN 2022 funded by the Italian Ministero dell’Universit\`a e della Ricerca (MUR) and by the European Union – Next Generation EU.  
This work is (partially) supported by ICSC – Centro Nazionale di Ricerca in High Performance Computing, Big Data and Quantum Computing, funded by European Union–NextGenerationEU.
O.S. and L.P. acknowledge partial financial support from
the INAF Minigrant 2023 \textit{Self-consistent Modeling of Interacting Binary Systems}
and O.S. acknowledges funding from the INAF large program \textit{BRAVOSUN}.
\end{acknowledgements}

%
   \bibliographystyle{aa} 
   \bibliography{axions.bib} 
%

\end{document}